\documentclass[twoside,epsfig]{article}
\usepackage{fleqn,espcrc2,epsfig}

\usepackage{graphicx}
\usepackage[figuresright]{rotating}

\newcommand{\AmS}{{\protect\the\textfont2
  A\kern-.1667em\lower.5ex\hbox{M}\kern-.125emS}}

\title{Two loop mass effects in the static position space QCD-potential}

\author{Michael Melles\address{Particle Theory\\
        Paul Scherrer Institute \\ 
        Villigen, CH-5232, Switzerland}}

\begin{document}
\begin{abstract}
The perturbatively calculable short distance QCD potential is known to two
loops including the effect of massive quarks. Recently, a simple
approximate solution in momentum space was utilized to obtain 
the potential in coordinate space. The latter is important in several respects.
A comparison with non-perturbative lattice results is feasible
in the overlap regime using light $\overline{\mbox{MS}}$ masses.
This might be even more promising employing the concept of the force
between the heavy color singlet sources, which can be easily derived from
the potential. 
In addition, the better than two percent accuracy 
bottom mass determination from $\Upsilon$-mesons 
is sensitive to massive charm loops at the two loop order.
We summarize recent results using exact one loop functions and explicit
decoupling parametrizations.
\end{abstract}

\maketitle

In analogy to QED, the heavy quark potential \cite{s,f} 
is of central interest in QCD
as a measure of the strong coupling. In the long distance regime $\sim {\cal O}$(1fm) 
it
contains the confining linear tail. With the inclusion of dynamical
quarks on the lattice, 
one expects to enter the so called ``string breaking'' regime at 
distances of ${\cal O}$(2fm) yielding the famous Yukawa potential \cite{y}.
At very short distances ($\leq$ 0.1fm)
one recovers the Coulomb-part of the potential,
modified by loop corrections like in QED. The overlap region is naturally
of considerable interest since non-perturbative effects might also enter \cite{b} and  
possibly as a ``normalization'' for lattice calculations. The important point
to notice here is that fermions with light $\overline{\mbox{MS}}$-masses
are easier to implement on the lattice.
For a direct comparison, due to its smaller value, the force \cite{g} between
the static sources is also of special interest. For a recent review of
lattice results on the forces in heavy bound states see Ref. \cite{b1}.  
From a phenomenological point of view, the importance of calculating mass
effects in the perturbative part of the 
heavy quark potential is two fold. On the one hand,
since we are dealing with a physical system, the corresponding physical
coupling definition (V-scheme) has several welcome advantages compared
to massless (calculational) schemes. $\alpha_V(Q^2/m^2)$ is an observable
i.e. the flavor thresholds are analytic, display automatic decoupling
of heavy quarks and are independent of the renormalization scale to the
order we are working. In addition, the physical scale of the problem is
determined by the transfered momentum.
Detailed discussions of the flavor threshold treatment
using analyitc schemes are contained in Refs. \cite{bgmr,m0,bmr}. 
On the other hand, the two loop mass corrections
to the heavy quark potential are important for the better than two percent
accuracy determination of the bottom mass \cite{hm}. In this case it is 
important not to treat the charm loops as massless but to consider the
full massive calculation.
Using the
physical $\Upsilon$-meson for this purpose, the effect of the mass shift
$\delta m_b$ depends on $\langle \phi_{1s} | V_F(r,m_c) | \phi_{1s}
\rangle$, where $\phi_{1s}=\frac{1}{\sqrt{\pi}} \left(
\frac{m_b C_F \alpha_s}{2} \right)^\frac{3}{2} \exp \left(- 
\frac{C_F \alpha_s}{2} m_b r \right)$ 
denotes the 1s ground state wave function of the
$\Upsilon$-meson and $V_F$ the massive fermionic corrections to the potential.
Note that in momentum space one would have an additional integration since
each wave function depends on a different three momentum making the overall
calculation prohibitive.

We begin, by recalling the definition of the potential through the
manifestly gauge invariant vacuum expectation value of the Wilson loop
$W_\Gamma=\langle 0 | \mbox{Tr}
{\cal P} \exp \left( i g \oint_\Gamma d x_\mu A^\mu_a {\mbox T}^a \right)
|0 \rangle $ of spatial extension
$r$ and large temporal extension $T$ with gluon exchanges between the
temporal lines.
The path-ordering ${\cal P}$ is necessary due to the non-commutativity of the SU(3)
generators $\mbox{T}^a$. In
the perturbative analysis through two loops considered here, all spatial
components of the gauge fields $A^\mu_a ({\bf r}, \pm T/2)$ can at most
depend on a power of $\log T$ and are thus negligible here. Furthermore,
$W_\Gamma \stackrel{T \to \infty}{\longrightarrow} \exp \left( - i T E_0 \right
)$, where the
ground state energy $E_0$ is identified with the potential $V$. 
Thus we arrive at the definition:
\begin{eqnarray}
&& V(r,m) = - \lim_{T \to \infty} \frac{1}{iT} \nonumber \\ &&
\log \langle 0| \mbox{Tr}
{\cal P} \exp \left( i g \oint_\Gamma d x_\mu A^\mu_a {\mbox T^a} \right)
|0 \rangle \label{eq:hqpdef}
\end{eqnarray}
Writing the source term of the heavy charges, separated at the distance
$r \equiv |{\bf r}-{\bf r}^\prime |$, as
\begin{equation}
J^a_\mu (x) = i g v_\mu T^a \left[ \delta ( {\bf x}
- {\bf r} )- \delta ( {\bf x}-{\bf r}^\prime ) \right]
\end{equation}
and neglecting contributions connecting the spatial components,
the perturbative potential is given by
\begin{eqnarray}
&& V(r,m)=-\lim_{T \to \infty} \frac{1}{iT}  \nonumber \\
&& \log \langle 0 | \mbox{Tr} {\cal T}
\exp \left( \int d^4x A^a_\mu (x)  J^\mu_a (x) \right) |0 \rangle \label{eq:vj}
\end{eqnarray}
In the above equation
$v_\mu=\delta_{\mu,0}$ due to the purely timelike nature of the sources. 
The potential in momentum space is the Fourier transform of $V(r)$. It can be
calculated directly in momentum space from the on-shell heavy quark anti-quark
scattering amplitude (divided by i) at the physical momentum transfer {\bf q},
projected onto the color singlet sector.
The potential can be used to define the effective charge $\alpha_V$ (
the so-called V-scheme)
through:
\begin{eqnarray}
V(Q,m) &\equiv& - 4 \pi C_F \frac{\alpha_V (Q,m)}{Q^2} \\ V(r,m)
&\equiv& - C_F \frac{\alpha_V(r,m)}{r}
\end{eqnarray}
where $Q^2 \equiv {\bf q}^2=-q^2$ and both expressions above are related
through a Fourier-transform, i.e. $V(r,m)= \int \frac{d^3Q}{(2 \pi)^3} V(Q,m)
\exp ( i \bf{Qr})$. At lowest order we obtain therefore the
well known Coulomb potential $V_C$ in each representation.

\begin{figure}[t] \vspace{-0.5cm}
\begin{center}
\epsfig{file=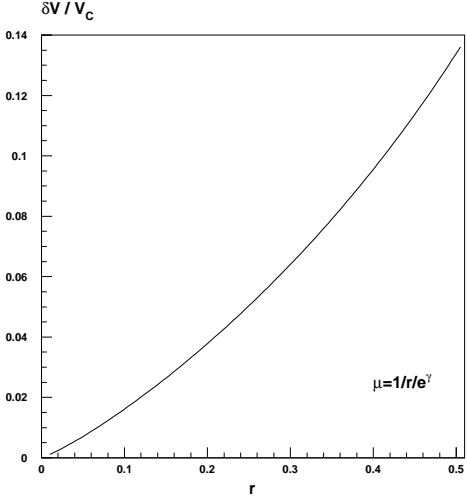,width=7.5cm}
\vspace{-1.5cm}
\caption{The size of the charm-mass corrections in Eq. \ref{eq:dVr} relative to
the Coulomb potential. The choice of the natural renormalization scale is
indicated in the figure. The distance $r$ between the sources is in GeV$^{-1}$.}
\label{fig:dvov}
\end{center}
\end{figure}

In Ref. \cite{m2}
the two loop corrections in momentum space were given in
approximate form based on the reconstructed solutions of the Gell-Mann Low
function with massive quarks obtained in Ref. \cite{bmr} from the exact numerical
results in Ref. \cite{m1}. In order to render decoupling of heavy flavors
with pole mass $m$ explicit one has to use the decoupling relation \cite{lrv}:
\begin{eqnarray}
&& \!\!\!\!\!\!\!\!\!\!\!\! \alpha_s^{(n_l)} (\mu) = 
\alpha_s^{(n_l-1)}(\mu) 
\! \left\{ 1 + \frac{
\alpha_s^{(n_l-1)}(\mu)}{\pi} \frac{1}{6} \log \frac{\mu^2}{m^2} + \right. \nonumber \\
&& \!\!\!\!\!\!\!\!\!\!\!\!\! \left. \left( \!\! \frac{\alpha_s^{(n_l-1)}(\mu)}{\pi} 
\!\! \right)^{\!\!2} \!\!\! \left[
\frac{7}{24}+\frac{19}{24} \log \frac{\mu^2}{m^2} +\frac{1}{36} \log^2
\frac{\mu^2}{m^2} \right] \!\! \right\} \label{eq:dec}
\end{eqnarray}
It is then useful to write the mass corrections to the potential in such a
way that the light quark with mass $m$ is included in the evolution of the
strong coupling. Thus the mass corrections vanish in the limits
$m \longrightarrow 0$ and $Q^2 \longrightarrow \infty$. We find from 
the results of Ref. \cite{m2} the following expression:
\begin{equation}
V^{\mbox{\tiny{NNL}}}(Q,m) = V^{\mbox{\tiny{NNL}}}(Q,0) + \delta V^{\mbox{\tiny{NNL}}}(Q,m)
\end{equation}
where the first term on the r.h.s. is given in Ref. \cite{ys} correcting an
error in the original calculation of the pure glue part of Ref. \cite{p1,jps}.
Denoting the
one loop subtracted mass correction function by
\begin{eqnarray}
&& \!\!\!\!\!\!\!\!\!\!\!\! \mbox{P} \left( \frac{m^2}{Q^2} \right) 
\equiv \frac{5}{3}-\log \frac{Q^2}{m^2} + 2 Q^2 \!\! \int^\infty_1 
\!\!\!\! \frac{dx \, f(x)}
{Q^2+4m^2x^2} 
\\ && \!\!\!\!\!\!\!\!\!\!\!\! \mbox{with} \nonumber \\
&& \!\!\!\!\!\!\!\!\!\!\!\! 
f(x) \equiv \sqrt{x^2-1} \frac{2x^2+1}{2x^4}
\end{eqnarray}
we find
\begin{eqnarray}
&& \!\!\!\!\!\!\!\!\!\!\!\! \delta V^{\mbox{\tiny{NNL}}}(Q,m)= V_C (Q^2)
\left\{
\frac{\alpha^{(n_l)}_{\overline{\mbox{\tiny{MS}}}}
(\mu)}{4 \pi} \frac{4}{3} T_F \mbox{P} \left( 
\frac{m^2}{Q^2} \right) \right. 
\nonumber \\ && \!\!\!\!\!\!\!\!\!\!\!\!\! 
\left( \frac{\alpha^{(n_l)}_{\overline{\mbox{\tiny{MS}}}}(\mu)}{4 \pi} \right)^2 \left[ \frac{8}{3} T_F 
\mbox{P} \left( \frac{m^2}{Q^2} \right) \left( \frac{31}{9} C_A-\frac{20}{9}
T_F n_l \right. \right. \nonumber \\ &&
\!\!\!\!\!\!\!\!\!\!\!\!\! \left. - \beta_0 \log \frac{Q^2}{\mu^2} \right)
+ \frac{16}{9}T_F^2 \mbox{P}^2 \left( \frac{m^2}{Q^2} \right) + \frac{76}{3}
T_F \times \nonumber \\
&& \!\!\!\!\!\!\!\!\!\!\!\!\! \left[ c_1 \! \int^\infty_{c_2} \!
\frac{dx}{x} \frac{2 Q^2}{Q^2+ 4 m^2 x^2} + d_1 \! \int^\infty_{d_2} \! 
\frac{dx}{x} \frac{2 Q^2}{Q^2+ 4 m^2 x^2} \right. 
\nonumber \\ 
&& \!\!\!\!\!\!\!\!\!\!\!\!\! \left. \left. + \frac{161}{114} +\frac{26}{19}
\zeta_3 -\log \frac{Q^2}{m^2} \right] \right\} \label{eq:dVQ}
\end{eqnarray}
The fitting constants are given by $c_2=0.47 \pm 0.01$, $d_2=1.12 \pm 0.02$
with $c_1,d_1$ fixed by the two conditions $c_1+d_1=1$ and
$c_1 \log ( 4 c_2^2)+ d_1 \log \log ( 4 d_2^2)= \frac{161}{114}+\frac{26}{19}
\zeta_3$ in order to ensure exact decoupling when using relation \ref{eq:dec}.
Eq. \ref{eq:dVQ} can be used to obtain the corresponding results in
coordinate space. Here we find analogously (with $\hat{m}=
e^{ \gamma } m$):
\begin{eqnarray}
&& \!\!\!\!\!\!\!\!\!\!\!\! \delta V^{\mbox{\tiny{NNL}}} (r,m)= V_C (r) \left\{
\frac{\alpha^{(n_l)}_{\overline{\mbox{\tiny{MS}}}}(\mu)}{3 \pi} \left[ \log ( \hat{m} r)
+ \frac{5}{6} + \right. \right.   \nonumber \\
&& \!\!\!\!\!\!\!\!\!\!\!\! \left. \int^\infty_1 \!\!\!\! dx \, e^{-2mrx}
f(x) \right] + \left( \!
\frac{\alpha^{(n_l)}_{\overline{\mbox{\tiny{MS}}}}(\mu)}{3 \pi} \! 
\right)^{\!\!2} \!\! \left[ -\frac{3}{2} \! \int^\infty_1 \!
\!\!\! dx \, f(x) \right. \nonumber \\ 
&& \!\!\!\!\!\!\!\!\!\!\!\! e^{-2 m rx} \!\! \left( \! \beta_0 \! \left( \! 
\log \! \frac{4m^2x^2}{\mu^2}
- \mbox{Ei}(2mrx)- \mbox{Ei}(-2mrx) \!\! \right) \right. \nonumber \\
&& \!\!\!\!\!\!\!\!\!\!\!\! \left. - \frac{31}{9} C_A+ \frac{20}{9}
T_F n_l \! \right) + \beta_0 \frac{\pi^2}{4} + 
3 \left( \! \log (\hat{m} r) + \frac{5}{6} \! \right) \!\! \times \nonumber \\
&& \!\!\!\!\!\!\!\!\!\!\!\! \left( \! \beta_0 \log ( \hat{\mu} r)
+ \frac{31}{18} C_A- \frac{10}{9} T_F n_l \! \right) \!\!- \!\!\! 
\int^\infty_1 \!\!\!\!\!\!\! dx
\, f(x) e^{-2mrx} \nonumber \\
&& \!\!\!\!\!\!\!\!\!\!\!\!
\left( \frac{1}{x^2} + x f(x) 
\log  \frac{x-  \sqrt{x^2-1}}{
x + \sqrt{x^2-1}} - \mbox{Ei} ( 2mxr) - \right. \nonumber \\
&& \!\!\!\!\!\!\!\!\!\!\!\!  \mbox{Ei} (-2mxr) + \log ( 4 x^2)
\Bigg) + \frac{\pi^2}{12} + \left( \! \log (\hat{m}r)+ \frac{5}{6} \right)^{\!\!2}
\nonumber \\ 
&& \!\!\!\!\!\!\!\!\!\!\!\! 
+ \frac{57}{4} \left(
\frac{161}{228}+\frac{13}{19} \zeta_3 + \log (\hat{m}r) + 
c_1 \mbox{Ei} (1,2c_2mr) \right. \nonumber \\
&& \!\!\!\!\!\!\!\!\!\!\!\!  +d_1 \mbox{Ei} (1,2d_2mr)
\Bigg) \Bigg] \Bigg\} \label{eq:dVr}
\end{eqnarray}
where $\mbox{Ei} (1,x) = \int^\infty_x \exp (-t) \frac{dt}{t}$. The relation
$\mbox{Ei}(x)+\mbox{Ei}(-x)= \mbox{P} \int^\infty_0 \exp [(1-t)x] \frac{2t \,
dt}{1-t^2}$ is also useful, with P denoting the principal value prescription.
While Eq. \ref{eq:dVr} vanishes for $m \longrightarrow 0$, it does not
for $r \longrightarrow 0$. It is a rather interesting fact that through
the Fourier transform of Eq. \ref{eq:dVQ} a non-analytic\footnote{Non-analytic
refers here to the fact that Eq. \ref{eq:dVrm} should be read as being proportional
to $\sqrt{m^2}$ since
$m^2$ is the only mass term entering the momentum space result in Eq. \ref{eq:dVQ}.}linear mass term
is generated which is furthermore enhanced by a factor $\pi^2$. It is
important to note that the linear mass term cannot be obtained by
first expanding Eq. \ref{eq:dVr} and then integrating over the dispersion
relation variable $x$, since it originates from large values of $x$.
This means it originates from momenta smaller than the quark mass, i.e.
it is of infrared origin. We find the following
limit:
\begin{eqnarray}
&& \!\!\!\!\!\!\!\!\!\!\!\! \delta V^{\mbox{\tiny{NNL}}} (r,m) \stackrel{r \ll
1/m}{\longrightarrow} -  m \frac{C_F}{4} 
\left( \alpha^{(n_l)}_{\overline{\mbox{\tiny{MS}}}}(\mu) \right)^2 \Bigg\{1+
\nonumber \\ && \!\!\!\!\!\!\!\!\!\!\!\!
\frac{\alpha^{(n_l)}_{\overline{\mbox{\tiny{MS}}}}(\mu)}{2 \pi} \! \Bigg[
\frac{31}{9} C_A- \frac{20}{9} T_F n_l + \beta_0 \! \Bigg( \!\! \log \frac{\mu^2}{m^2}-
4 \log (2) \nonumber \\ && \!\!\!\!\!\!\!\!\!\!\!\!
+ 3 \Bigg) + \frac{4}{45} \left( 31 - 30 \log (2) \right) +
\frac{76}{3 \pi} \left( c_1 c_2 + d_1 d_2 \right) \Bigg] \Bigg\} 
\nonumber \\ && \!\!\!\!\!\!\!\!\!\!\!\!
+ {\cal O}
(m^2 r) \label{eq:dVrm}
\end{eqnarray}
In terms of the $\overline{\mbox{MS}}$-mass parameter $ \overline{m}(\mu)$, 
one only needs to
use the relation 
\begin{equation}
m=\overline{m}(\mu) \! \left[ 1 + C_F \frac{ 
\alpha_{\overline{\mbox{\tiny MS}}} 
(\mu)}{ \pi} \!\! \left( 1 + \frac{3}{2} \log \frac{\mu}{ \overline{m}(\mu)} \right) \right]
\end{equation}
in the one loop term to obtain the NNL correction.
The size of the mass corrections relative to the Born Coulomb potential
is displayed in Fig. \ref{fig:dvov} for the charm quark mass $m=1.5$GeV and
the ``natural'' renormalization scale choice $\mu=\frac{1}{r e^\gamma}$, since
the $\gamma$ terms are generated by the Fourier transform. In Ref. \cite{p}
it is shown that this scale choice is almost identical to the BLM-scale \cite{blm}
and thus consistently absorbs large renormalization group logarithms.
The two loop running coupling is given by
\begin{equation}
\alpha_{\overline{\mbox{\tiny MS}}} (\mu)\! = \! \frac{4 \pi}{\beta_0 \log
\frac{\mu^2}{\Lambda^2_{QCD}}} \!\!
\left( \!\! 1 \! - \! \frac{\beta_1}{\beta_0^2}
\frac{ \log \left( \log \frac{\mu^2}{\Lambda^2_{QCD}} \right)}{\log
\frac{\mu^2}{\Lambda^2_{QCD}}} \!\! \right) \label{eq:Lqcd}
\end{equation}
where we normalize the QCD-scale parameter $\Lambda_{QCD}$ such that
$\alpha_{\overline{\mbox{\tiny MS}}} (M_Z) = 0.12$ which corresponds to
$\Lambda_{QCD}=0.25$GeV and we keep $n_l=4$ fixed.
The first two terms of the $\beta$-function are 
gauge invariant and scheme independent in
massless renormalization schemes and are given by 
$\beta_0= \frac{11}{3} C_A - \frac{4}{3} T_F n_l$ and $\beta_1
= \frac{34}{3} C_A^2 - \frac{20}{3}C_A T_F n_l - 4 C_F T_F n_l$. In QCD we have
$C_A=3$, $C_F= \frac{4}{3}$ and $T_F=\frac{1}{2}$.
The effect is at the several percent level and increases above 10 \% for
distances of 0.1 fm ($ \sim $ 0.5 GeV$^{-1}$). It vanishes for small $r$
since we are displaying the corrections relative to $V_C$. In absolute
terms we checked numerically that Eq. \ref{eq:dVrm} is reproduced for
small distances by the full result in Eq. \ref{eq:dVr}.
The inclusion of these charm quark
corrections into a full sum rule determination
of the bottom quark mass in Ref. \cite{h} led to a shift of $-30$ MeV and
yields $\overline{m}_b(\overline{m}_b)=4.17 \pm 0.05$ GeV, which in light
of the small error is a significant contribution.

Instead of considering the potential, or $\alpha_V(r,m)$, it is also of
interest to study the coupling $\alpha_F(r,m)=-r^2 
\frac{\partial (\alpha_V(r,m)/r)}{\partial r}$ which is defined from the
force between the static quarks. In general $\alpha_F(r,m)$ is smaller
than $\alpha_V(r,m)$ or even $\alpha_{\overline{\mbox{\tiny{MS}}}}(\mu=1/r)$
\cite{m2}, which makes it suitable as an expansion parameter in perturbative
calculations. It is also useful in lattice calculations to determine
$\alpha_s$ \cite{m,rs}.
The force is given by $F(r,m)=-\frac{\partial V(r,m)}{\partial r}$
and the massless case is taken from Ref. \cite{m2}: 
\begin{eqnarray}
&& \!\!\!\!\!\!\!\!\!\!\!\! F^{\mbox{\tiny{NNL}}}(r,0) = -C_F \frac{
\alpha^{(n_l)}_{\overline{\mbox{\tiny{MS}}}}(\mu)}{r^2} \Bigg\{ 1 + \frac{
\alpha^{(n_l)}_{\overline{\mbox{\tiny{MS}}}}(\mu)}{4 \pi}  \Bigg( 
\nonumber \\ && \!\!\!\!\!\!\!\!\!\!\!\!
2 \beta_0
\log (\hat{\mu} r) 
+ f_1 \Bigg) + \left( \frac{
\alpha^{(n_l)}_{\overline{\mbox{\tiny{MS}}}}(\mu)}{4 \pi} \right)^2 \Bigg(
4 \beta_0^2 \log^2 (\hat{\mu} r) \nonumber \\ && \!\!\!\!\!\!\!\!\!\!\!\!
+ 2 \Big[ \beta_1+ 
2 \beta_0 f_1
 \Big] \log (\hat{\mu} r) 
+ f_2 \Bigg) \Bigg\}
\end{eqnarray}
where
\begin{eqnarray}
&& \!\!\!\!\!\!\!\!\!\!\!\!
f_1=-\frac{35}{9} C_A + \frac{4}{9} T_F n_l \label{eq:f10} \\
&& \!\!\!\!\!\!\!\!\!\!\!\!
f_2= \left(- \frac{7513}{162}+\frac{229}{27} \pi^2 - \frac{1}{
4} \pi^4 + \frac{22}{3} \zeta_3 \right) C_A^2 + \nonumber \\
&& \!\!\!\!\!\!\!\!\!\!\!\!
\left( \frac{3410}{81}-
\frac{88}{27} \pi^2-\frac{56}{3} 
\zeta_3 \right) C_AT_Fn_l - 
\Bigg( \frac{31}{3}-  \nonumber \\
&& \!\!\!\!\!\!\!\!\!\!\!\! 16 \zeta_3 \Bigg) C_FT_Fn_l - \left(
\frac{560}{81}-\frac{16}{27} \pi^2 \right) \left( T_Fn_l \right)^2 
\label{eq:f20}
\end{eqnarray}
From Eq. \ref{eq:dVr} we
thus find for the mass corrections the following expression:
\begin{eqnarray}
&& \!\!\!\!\!\!\!\!\!\!\!\!
\delta F^{\mbox{\tiny{NNL}}}(r,m)=-\frac{\partial \log V_C(r)}{\partial r} \delta 
V^{\mbox{\tiny{NNL}}} (r,m) - V_C \Bigg\{ \nonumber \\ 
&& \!\!\!\!\!\!\!\!\!\!\!\!
\frac{\alpha^{(n_l)}_{\overline{\mbox{\tiny{MS}}}}(\mu)}{3 \pi}  \left[ 
\frac{1}{r} - \!\! \int^\infty_1 \!\!\!\!\!\! dx \, 2mx f(x) e^{-2mrx} \right] 
\!\! + \!\!
\left( \!\! \frac{\alpha^{(n_l)}_{\overline{\mbox{\tiny{MS}}}}(\mu)}{3 \pi}
\!\! \right)^{\!\! 2} 
 \nonumber \\ && \!\!\!\!\!\!\!\!\!\!\!\!
\times \Bigg[ \frac{3}{2} \int^\infty_1  \!\!\!\! dx \, e^{-2mrx} f(x)
\left( 2 m x \Bigg( \beta_0 \Bigg[ \log \frac{4m^2x^2}{\mu^2} - \right. \nonumber \\ 
&& \!\!\!\!\!\!\!\!\!\!\!\! 
\mbox{Ei}(2mrx)
-\mbox{Ei}(-2mrx)  \Bigg] - \frac{31}{9}C_A+\frac{20}{9}T_Fn_l \Bigg)+ 
\nonumber
\\ && \!\!\!\!\!\!\!\!\!\!\!\!  \frac{\beta_0}{r} \left( e^{2mrx}+e^{-2mrx} \right) 
\!\! \Bigg) + \frac{3}{r} \beta_0 \left[ \log (\hat{m} r) 
+ \frac{5}{6} \right] \nonumber \\
&& \!\!\!\!\!\!\!\!\!\!\!\! + \frac{3}{r} \left( \beta_0 
\log (\hat{\mu} r) 
+ \frac{31}{18} C_A - \frac{10}{9} T_F n_l \right) + \nonumber \\
&& \!\!\!\!\!\!\!\!\!\!\!\! \int^\infty_1 \!\!\!\! dx \, e^{-2mrx} f(x) \Bigg( 
2 m x \Bigg( \log (4x^2) - \mbox{Ei}(2mrx)- \nonumber \\
&& \!\!\!\!\!\!\!\!\!\!\!\! \mbox{Ei}(-2mrx)+
\frac{1}{x^2} + x f(x) \log \frac{x-\sqrt{x^2-1}}{x+\sqrt{x^2-1}} 
\Bigg) + \nonumber \\ 
&& \!\!\!\!\!\!\!\!\!\!\!\! 
\frac{1}{r} \left( e^{2mrx}+e^{-2mrx} \right) \Bigg) + \frac{2}{r} \left(
\log (\hat{m}r) + \frac{5}{6} \right) + \nonumber \\
&& \!\!\!\!\!\!\!\!\!\!\!\!
\frac{57}{4r} \left( 1 - c_1 e^{-2c_2mr}-d_1 e^{-2d_2mr} \right) \Bigg] \Bigg\}
\label{eq:dFr}
\end{eqnarray}
\begin{figure} \vspace{-0.5cm}
\begin{center}
\epsfig{file=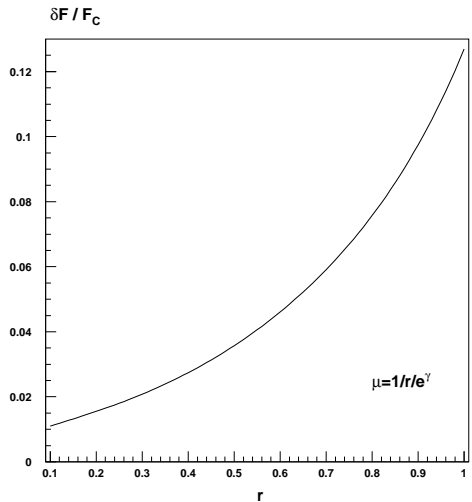,width=7.5cm}
\vspace{-1.5cm}
\caption{The size of the charm-mass corrections in Eq. \ref{eq:dFr} relative to
the lowest order force from the Coulomb potential. 
The choice of the natural renormalization scale is
indicated in the figure.
The distance $r$ between the sources is in GeV$^{-1}$.}
\label{fig:dfof}
\end{center}
\end{figure}
The size of the charm-mass corrections relative to the Born term is presented in
Fig. \ref{fig:dfof} for the same scale choice as in Fig. \ref{fig:dvov}.
It can be seen that the effect is less than half of that for the mass
corrections to the potential and start to increase more rapidly at distances
over 0.1 fm.

In summary, we have calculated the massive quark corrections to the static
QCD potential in coordinate space at the two loop level. 
The results presented here use the exact
one loop vacuum polarization functions and a dispersion relation fit based
on the results of Ref. \cite{m2}. The uncertainty is estimated at the percentile
level from comparisons with the exact Monte Carlo results in momentum space
of Ref. \cite{m1}. For the bottom mass determination, the inclusion of a massive
charm quark in the $\Upsilon$ potential
is significant due to two reasons. One factor is that the effective physical scale
depends parametrically on the charm mass, leading to a large value of 
the coupling. The other reason originates from the fact that 
in the relation between the potential contribution to
the static energy and the pole mass,
there is an uncanceled non-analytic linear mass term \cite{hm,h}, 
whose origin is the Fourier transform. Together, these 
two effects lead to a shift of
$-30$ MeV and
$\overline{m}_b(\overline{m}_b)=4.17 \pm 0.05$ GeV.
In addition we have calculated the mass corrections for the force between 
two static color charges in a singlet state at the two loop level.
The size of the effect in general is smaller than for the potential but
still significant at larger distances. At small distances, the linear
$r$-independent mass term from the potential drops out and thus leads
to smaller corrections in this regime.

\end{document}